\documentclass[runningheads]{llncs}
\usepackage[T1]{fontenc}
\usepackage{graphicx}
\usepackage{float}

\usepackage{booktabs}
\usepackage{tabularx}
\usepackage[table]{xcolor}

\usepackage[english]{babel}
\usepackage[autostyle, english=american]{csquotes}
\MakeOuterQuote{"}

\usepackage[hidelinks]{hyperref}

\usepackage{enumitem}

\usepackage[most]{tcolorbox}
\tcbuselibrary{breakable}
\tcbset{%any default parameters
  width=\linewidth,
  boxrule=-0pt,
  frame hidden,
  sharp corners,
  enhanced,
  notitle,
  left=0pt,top=0pt,bottom=0pt,right=0pt
}

\usepackage{cleveref}

\usepackage[subtle,title=normal,tracking=normal]{savetrees}

\begin{document}
\title{Towards an Architectural \textit{Perspective} for Sustainability: Bundle the Needs from Industry}
\author{Markus Funke\inst{1}\orcidID{0000-0003-2302-2555} \and
Patricia Lago\inst{1}\orcidID{0000-0002-2234-0845}}
\authorrunning{M. Funke and P. Lago}
\institute{Vrije Universiteit Amsterdam, The Netherlands\\
\email{\{m.t.funke,p.lago\}@vu.nl}}
\maketitle              % typeset the header of the contribution
\begin{abstract}
Sustainability is increasingly recognized as an emerging quality property in software-intensive systems, yet architects lack structured guidance to address it effectively throughout the software design phase.
Architectural perspectives---an architectural knowledge artifact composed of concerns, activities, tactics, pitfalls, and checklists---offer a promising approach to tackle such emerging quality properties across architectural views and are also independent of architecture frameworks and industry contexts.
In this paper, we present a \textit{sustainability perspective vision}, i.e., a revised notion of architectural perspective meant to be filled with its own elements to target sustainability concerns.
We formulate our sustainability perspective vision through evidence from applying snowballing to seminal literature and from conducting a focus group with experts in the field.
Our findings confirm the relevance of the different perspective elements in practice and highlight implications for shaping a sustainability perspective that meets industrial needs.

\end{abstract}
%
%
%
%%%%%%%%%%%%%%%%%%%%%%
%%%%% CONTENT
%%%%%%%%%%%%%%%%%%%%%%
%%%%%%%%%%%%%%%%%%%%%
%%% Introduction
%%%%%%%%%%%%%%%%%%%%%
\section{Introduction}
\label{sec:introduction}

Although sustainability is increasingly prioritized in the software industry and acknowledged as an essential quality property and architectural concern, guidance on how to embrace and tackle these concerns at the earliest stage of software design remains limited~\cite{HeldalEtAl_SustainabilityCompetencies_2024,FunkeLago_CarvingSustainability_2023}.
The challenge emerges as addressing sustainability related concerns are not yet embedded in traditional architecture design and decision-making processes. Architecture decisions, however, rely on experience and tacit knowledge~\cite{AliBabarEtAl_SoftwareArchitecture_2009,CapillaEtAl_10Years_2016}. Even though sustainability has gained prominence in recent years \cite{CaleroEtAl_5WsGreen_2020}, this necessary experience is not yet fully established.
Consequently, architects face uncertainty when translating sustainability concerns into concrete decisions which underscores the need for structured guidelines and supportive knowledge bases~\cite{HeldalEtAl_SustainabilityCompetencies_2024,FunkeLago_CarvingSustainability_2023,PathaniaEtAl_KnowledgeBase_2023}. Architectural perspectives \cite{WoodsRozanski_UsingArchitectural_2005}, in short \textit{perspectives}, could provide this systematic guidance.

The literature defines perspectives as \textit{"a collection of architectural activities, tactics, and guidelines that are used to ensure that a system exhibits a particular set of related quality properties that require consideration across a number of the system’s architectural views"} \cite[p.47]{RozanskiWoods_SoftwareSystems_2012}.

%% same
Perspectives are closely related to architecture viewpoints and views \cite{ISO_42010_2022}, aiming to address quality properties and architecture concerns. Viewpoints guide the architect in illustrating one or more quality concerns as views (e.g., diagrams or text), using recurring patterns and conventions \cite{ISO_42010_2022,RozanskiWoods_SoftwareSystems_2012}. For instance, the deployment viewpoint helps creating views for network and runtime concerns \cite{RozanskiWoods_SoftwareSystems_2012}. While viewpoints work on isolated concerns, cross-cutting concerns like security affect the entire architecture and multiple stakeholders \cite{RozanskiWoods_SoftwareSystems_2012,BouckeHolvoet_RelatingArchitectural_2006}, and must be addressed across multiple views.
Sustainability is one such cross-cutting concern, given its multifaceted nature \cite{McGuireEtAl_SustainabilityStratified_2023} and multi\-di\-men\-sional quality requirements \cite{LagoEtAl_FramingSustainability_2015}. Architectural perspectives offer structured guidance for addressing such concerns across multiple views. This adaptability makes them particularly relevant in industry, where systems must satisfy diverse and evolving requirements.

From an educational point of view, the need for a sustainability perspective is confirmed by our experience in teaching various Computer Science Master's courses at the Vrije Universiteit Amsterdam. 
Despite integrating sustainability into our curriculum for over a decade, students are still challenged with incorporating sustainability concerns into their architectural views due to a lack of available, generalized practices.
A well-defined sustainability perspective would equip future software engineers and architects with a structured framework, including tactics and guidelines, to better address sustainability in architecture and practice.

We believe a sustainability perspective can be a powerful tool in architecture design---if grounded in industry needs. As a knowledge base, it can support architects unfamiliar with sustainability and help businesses meet emerging requirements. Therefore, in this paper, we present a \textit{sustainability perspective vision} that guides architects in addressing sustainability concerns in software architecture views. We formulate and then illustrate the feasibility of our vision through evidence coming from the literature and experts in the field.

In \Cref{sec:relatedwork}, we discuss related work. \Cref{sec:method} outlines our research approach for designing a sustainability perspective. \Cref{sec:snowballing} positions perspectives in the literature via forward snowballing, while \Cref{sec:focus-group} presents focus group insights. \Cref{sec:vision} integrates the results to frame our envisioned perspective. \Cref{sec:threats} covers potential threats to validity, and \Cref{sec:conclusion} concludes with future work.

%%%%%%%%%%%%%%%%%%%%%
%%% Related work
%%%%%%%%%%%%%%%%%%%%%
\section{Related Work}
\label{sec:relatedwork}
In this section, we discuss concepts \textit{related} to the idea of (a) integrating sustainability into the software development life cycle, (b) addressing sustainability as cross-cutting concerns in architecture views, and (c) using perspectives as knowledge base to educate architects for such concerns. Literature focusing particularly on the notion of \textit{perspectives} according to Woods and Rozanski \cite{WoodsRozanski_UsingArchitectural_2005} are systematically analyzed in \Cref{sec:snowballing} by performing snowballing.

\paragraph{Software Development Life Cycle.}
The software development life cycle (SDLC) \cite{Ruparelia_SoftwareDevelopment_2010}, and thus research on integrating sustainability, spans different stages.
At the \textbf{requirements} stage, studies aim to elicit sustainability concerns early \cite{BeckerEtAl_SustainabilityDesign_2015,ChitchyanEtAl_SustainabilityDesign_2016,MahauxEtAl_DiscoveringSustainability_2011}.
During \textbf{design}, studies support architects in in their design and decision making \cite{AndrikopoulosEtAl_SustainabilitySoftware_2022,LagoEtAl_SustainabilityAssessmentToolkit_2024,KoziolekEtAl_MORPHOSISLightweight_2012a,VolpatoEtAl_HasSocial_2019}.
Software architecture tactics (which are one element of perspectives) can improve energy efficiency \cite{VosEtAl_ArchitecturalTactics_2022,FunkeEtAl_ExperimentalEvaluation_2024}, but typically target single quality attributes.
At \textbf{development} time, work focuses on green coding practices and energy-efficient languages \cite{RaniEtAl_EnergyPatterns_2024,PereiraEtAl_RankingProgramming_2021}.
For \textbf{maintenance}, runtime sustainability monitoring has been explored \cite{GuldnerEtAl_DevelopmentEvaluation_2024,DingaEtAl_EmpiricalEvaluation_2023}.

Most existing contributions are isolated, lack integration, or are not aligned with industry practice due to limited involvement. Our research seeks two goals: (i) combine and reuse existing tools, methods, and frameworks to target sustainability at design time and create a holistic knowledge base; and (ii) involve experts upfront to identify which sustainability concerns require guidance and knowledge.

\paragraph{Cross-cutting concerns.}
Independently from perspectives, research recognized the problem of addressing concerns in multiple viewpoints as challenging. 
Boucké and Holvoet~\cite{BouckeHolvoet_RelatingArchitectural_2006} describe that the misalignment between views and drivers can lead to \textit{tangling}, i.e., the confusion among architects introduced by addressing multiple architectural concerns by one single view. To overcome this, the authors argue for extending the architectural description through "slices" and "connectors" to make relations between views more explicit. 
While this approach focuses particularly on tracking relationships between views, it does not \textit{guide} the architect in creating such views for a specific concern. In contrast, perspectives propose for each individual concern related activities. 

Both the cross-cutting views from Boucké and Holvoet \cite{BouckeHolvoet_RelatingArchitectural_2006} and the perspectives from Woods and Rozanski \cite{WoodsRozanski_UsingArchitectural_2005}, recognize relations and similarities to aspect-oriented programming (AOP) \cite{KiczalesEtAl_AspectorientedProgramming_1997}. However, while AOP intends to improve re-usability, maintainability, and scalability at code level, we need to integrate the cross-cutting nature of sustainability at design time. For a complete comparison between perspectives and AOP, we suggest the original work from Woods and Rozanski~\cite{WoodsRozanski_UsingArchitectural_2005}.

\paragraph{Knowledge base.}
Pathania et al. \cite{PathaniaEtAl_KnowledgeBase_2023} share our line of reasoning, emphasizing the lack of a standard knowledge base for software development regarding sus\-tain\-ability---but at implementation level. In their paper, the authors envision a knowledge base for common
"sustainability weaknesses" at code level (i.e., "part
of code that has a detrimental impact on energy consumption" \cite{PathaniaEtAl_KnowledgeBase_2023}). The authors propose to re-evaluate existing weaknesses emerged in other disciplines and standards, such as security or performance, and classify their impact on sustainability.

To the best of our knowledge, no comprehensive knowledge base for architecture exists, targeting emerging sustainability concerns like energy consumption, resource utilization, scalability, or reusability.
Therefore, the development of a method that addresses both, the cross-cutting nature of sustainability in architecture views, and the lack of guidance to tackle sustainability concerns is the key position of our research.

%%%%%%%%%%%%%%%%%%%%%
%%% Method
%%%%%%%%%%%%%%%%%%%%%
\section{The Envisioned Approach}
\label{sec:method}
\begin{figure*}[htbp!]
    \centering
    \includegraphics[width=0.9\textwidth]{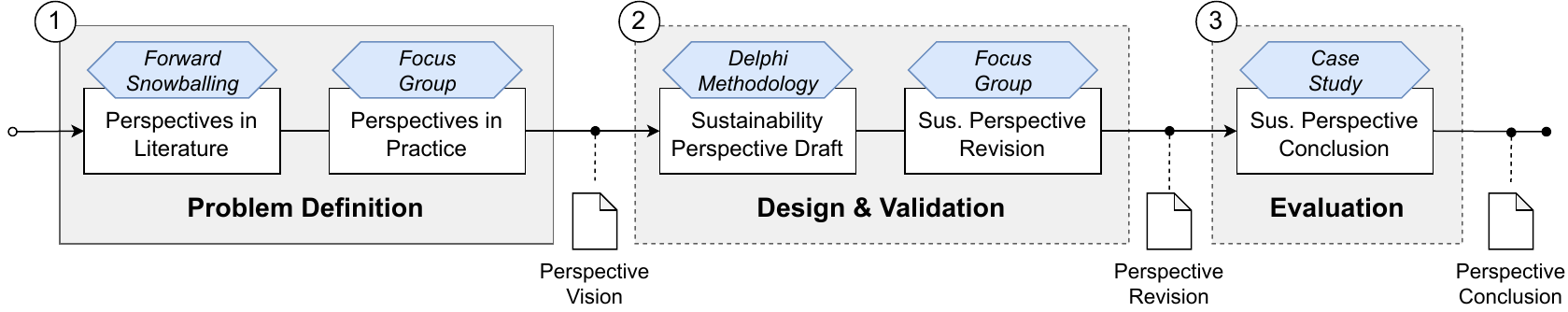}
    \caption{Research approach adopted from \cite{poster_FunkeLago_2024}. Phase 1 is this current work; phase 2 and 3 are planned and have yet to be executed.} 
    \label{fig:study-design}
\end{figure*}

To propose a sustainability perspective which is both grounded in theory and effective for its usage in industrial practice we follow the research approach and different Design Science Research (DSR) \cite{Wieringa_DesignScience_2014} phases put forward in \cite{poster_FunkeLago_2024}. As illustrated in \Cref{fig:study-design}, this present paper carries out the first iteration of the DSR cycle, i.e., Phase 1, and frames the perspective vision accordingly.
Below, the research design of Phase 1 is described in detail.
The other phases are described in \cite{poster_FunkeLago_2024} and remain for future work. 
In this present paper, we first review the literature on the notion of architecture perspectives to build a theoretical foundation (snowballing). Then, we supplement these findings by talking to experts, informing us on the relevance and usage of perspectives in industry (focus group). Through both, we are able to describe our first artifact---the perspective vision framed by literature and practice.

\subsection*{Phase 1: Problem Definition (this paper)}

\textbf{Snowballing.} We adopt a forward snowballing approach to systematically scan the literature for existing perspectives potentially related to our vision. Rozanski and Woods \cite{RozanskiWoods_SoftwareSystems_2012} propose a perspective catalog, defining a set of perspectives based on their experience (e.g., a security perspective, or a performance and scalability perspective). Scanning the literature for new perspectives would allow us to understand how this catalog has evolved over time. Literature developing novel perspectives might also help us in crafting a new one by studying their applied methodology, gained experience, and potential pitfalls to consider.

Forward snowballing involves searching for work that cite previously identified and relevant studies. We follow the guidelines for snowballing described by Wohlin~\cite{Wohlin_GuidelinesSnowballing_2014}. While snowballing has been explored as a complementary approach to traditional database searches, forward snowballing has also been used as valuable alternative to source primary studies (e.g., \cite{VentersEtAl_SustainableSoftware_2023,LwakatareEtAl_LargescaleMachine_2020}). Felizardo et al.~\cite{FelizardoEtAl_UsingForward_2016} demonstrate that using forward snowballing can lead to higher precision and slightly lower recall compared to database searches. However, the risk of potentially overlooking relevant studies remains \cite{WohlinEtAl_SuccessfulCombination_2022,FelizardoEtAl_UsingForward_2016}. To mitigate this, we conduct one round of backward snowballing, i.e., analyzing the reference list of our primary studies, to identify potential other relevant studies.

We \textit{consciously} focus exclusively on literature discussing the notion of perspectives according to Woods and Rozanski (see \Cref{sec:snowballing} for all inclusion and exclusion criteria). While individual concepts---such as quality concerns and tactics---are indeed applied in isolation during the architecture lifecycle, perspectives integrate all these elements to one cohesive knowledge base. For emerging concerns like sustainability, where knowledge is still limited or scattered, such knowledge base is key in education and supporting architects during the design phase. Therefore, we do not conduct a systematic literature review (SLR) on, e.g., energy efficiency tactics or sustainability-related frameworks. This approach would not address our central question in the problem definition phase, of how a sustainability perspective needs to be designed to meet the needs of industrial experts.

We base our starting set ("seed set" \cite{FelizardoEtAl_UsingForward_2016}) on the two earliest works in which the notion of architectural perspectives was first proposed: (i) the research paper by Woods and Rozanski \cite{WoodsRozanski_UsingArchitectural_2005}, and (ii) the book by Rozanski and Woods \cite{RozanskiWoods_SoftwareSystems_2012} that followed the paper a year later. Papers relating to perspectives should cite at least one of these works. Further details and results are in \Cref{sec:snowballing}.
\\
\\
\noindent
\textbf{Focus Group.} 
Since the notion of perspectives was already proposed 20 years ago at the time of writing, we want to determine industrial relevance of perspectives and whether the concept has evolved with industry needs. This understanding will allow us to elicit the requirements to deliver an artifact that is useful for professional practice. To this end, we conduct a focus group. 

We apply the guidelines for focus groups according to Kontio et al. \cite{KontioEtAl_FocusGroup_2008}. We use purposive sampling to select our members, focusing on their individual characteristics and our personal established connections. We decided to select two groups of experts: group (i) is familiar with the original notion of perspectives; while group (ii) is not. Instead of interviewing both groups in isolation, our setting enables the members of both groups to build on each others experiences, increasing richness and information on the topic at hand, which is in accordance with the guidelines of \cite{KontioEtAl_FocusGroup_2008}. The concrete focus group setup and results are outlined in \Cref{sec:focus-group}.

The review of perspectives in the literature and in practice form the basis of our problem definition and lay the foundation for our research. We use the information gained as the requirements for designing and building our future artifact.

%%%%%%%%%%%%%%%%%%%%%
%%% Related Work
%%%%%%%%%%%%%%%%%%%%%
\section{Results - \textit{Perspectives} in Literature}
\label{sec:snowballing}

\textbf{Setup.} To mitigate the risk of missing any related (scientific) work on creating novel architectural perspectives, we performed forward snowballing based on the meta-search engine Google Scholar. Research showed that Google Scholar has the highest intersection compared to Web of Science and Scopus \cite{Martin-MartinEtAl_GoogleScholar_2018}; and it is not restricted to one database providing a robust way retrieving the necessary citations for forward snowballing.

Our inclusion criteria were 
(\textbf{IC\textsubscript{1}}) Study cites either \cite{WoodsRozanski_UsingArchitectural_2005} or \cite{RozanskiWoods_SoftwareSystems_2012}; and
(\textbf{IC\textsubscript{2}}) Study develops a novel architectural perspective.
We excluded studies based on the following criteria:
(\textbf{EC\textsubscript{1}}) Duplication of an already considered study;
(\textbf{EC\textsubscript{2}}) Study not in English; or
(\textbf{EC\textsubscript{3}}) Study not accessible.

In total, we gathered 1000 citations\footnote{\scriptsize Google Scholar constraints the result set to the first 1000 relevant results (see: \url{https://scholar.google.com/intl/en/scholar/help.html\#export}).} for the book \cite{RozanskiWoods_SoftwareSystems_2012} (executed on 23-07-2024) and 41 citations for the paper \cite{WoodsRozanski_UsingArchitectural_2005} (executed on 03-03-2025). To obtain a manageable number of studies for the 1000 book results, we further limited the result set to the keyword `perspective' present in the title of the study. This lowered the result set to 16 studies. We further validated relevance by reading the abstract, followed by the full-text. After filtering, we obtained 6 papers to be considered as primary studies. To mitigate the risk of overlooking relevant literature, we conducted one round of backward snowballing on the references of our 6 primary studies; however, this did not yield any additional studies. For transparency, the complete snowballing process is available in our online supplementary material~\cite{replication_package}. Below, we synthesize the results.
\newline
\newline
\noindent
\textbf{Synthesis.} Some research developed novel perspectives for quality concerns which were not yet part of the perspectives catalog \cite{RozanskiWoods_SoftwareSystems_2012}: scalability \cite{TekinerdoganOzcan_ArchitecturalPerspective_2017}; safety \cite{GurbuzEtAl_SafetyPerspective_2014}; and learning analytics \cite{KiwelekarEtAl_ArchitecturalPerspective_2020}. The scalability and safety perspectives follow the perspective concept as proposed strictly. The learning analytics perspective, however, deviates. 
Instead of \textit{applying} their perspective on an existing architecture and its views, or \textit{considering} it during system design to challenge decisions, the authors \textit{create} a perspective which can be considered as an architectural view and hence as output of the system design. This "misinterpretation" will be discussed in \Cref{sec:focus-group}.

One research \cite{KanelVecchiola_GlobalTechnology_2013} adopts the existing `Availability and Resilience' perspective from the catalog. They utilize perspectives to explore and provide principles for resilient information systems, such as decision support platforms. This approach addresses the issue from an architectural design standpoint, incorporating resilience principles from existing literature. Consequently, the architectural perspective proposed by Woods et al. aids in analyzing current architectures to ensure their resilience in emergency management situations.

Sant'Anna et al. \cite{SantAnnaEtAl_MasteringCrosscutting_2013} do not propose a specific perspective. However, the authors address the problem of cross-cutting concerns also from an architectural documentation approach, i.e., with "aspects", by proposing "aspectual templates". Such templates do also record architectural decisions and can be seen as complementary to perspective. We consider these as an interesting approach which might help us while designing our own, industry relevant perspective.

\color{black}
The work from Jagroep et al. \cite{JagroepEtAl_ExtendingSoftware_2017} can be considered as the most related to ours. The authors propose an energy consumption perspective and validate it with a case study. The perspective is based on a new quality attribute `sustainability' with its sub-characteristic `resource consumption'. 
The authors provide a set of measures and metrics for the identified quality properties `software utilisation', `workload energy', and `energy usage'. 
Even though such measures and metrics are indeed helpful as "response measures" that help evaluate how a software system reacts to certain conditions (stimuli) \cite{BassEtAl_SoftwareArchitecture_2021}, the concept of measures and metrics are not per se part of an architectural perspective as these do not help in targeting certain concerns \textit{within} an architecture view.
Considering the architecture lifecycle~\cite{TangEtAl_ComparativeStudy_2010}, we would assign such process of monitoring "energy hotspots" \cite{ProcacciantiEtAl_GreenLab_2015} rather to the architecture maintenance phase in order to improve and evaluate an already implemented architecture. During architecture synthesis (design), a perspective would serve as "general knowledge" that can be "applied" to solve the problem \cite{TangEtAl_ComparativeStudy_2010}.

Although this energy consumption perspective does not fully align with our broader vision of sustainability and its intended use on architecture views, we did not want to miss out on any relevant studies building on this work from Jagroep et al. \cite{JagroepEtAl_ExtendingSoftware_2017}. Therefore, we conducted one additional round of forward snowballing based on this work. The setup was identical with our initial snowballing round. In total, we gathered 49 citations. After screening titles, abstracts, and full-texts, we can conclude that none of the 49 papers are relevant for our sustainability perspective.

From our literature review we can now derive the following conclusions: 
(i) the catalog from Rozanski and Woods \cite{RozanskiWoods_SoftwareSystems_2012} could be indeed extended and updated through the identified perspectives, although they do not provide useful insights on how to \textit{develop} new perspectives; 
(ii) existing literature does not examine whether and how perspectives are used in industrial contexts; 
(iii) we found one promising perspective targeting energy consumption. However, in our vision of a sustainability perspective, we want to tackle the problem from an architecture knowledge angle rather than from a measurement angle. Nevertheless, we want to build up on the results of Jagroep et al. \cite{JagroepEtAl_ExtendingSoftware_2017} and target the existing gaps: (i) providing a \textit{complete} perspective including tactics, pitfalls, and the checklist; (ii) putting the work into a fresh light by updating it with a recent view sustainability; and (iii) reviewing the concept of perspectives by aligning them with the needs coming from industry.
\color{black}

%%%%%%%%%%%%%%%%%%%%%
%%% Focus Group
%%%%%%%%%%%%%%%%%%%%%
\section{Results - \textit{Perspectives} in Practice}
\label{sec:focus-group}

\textbf{Setup.} We conducted a focus group on 30-09-2024 to assess how perspectives, introduced in 2005, are adopted in industry and whether they have evolved with industrial needs. Following the typical focus group size (4-9 experts \cite{KontioEtAl_FocusGroup_2008}), we invited 4 participants (P1-P4). As described in \Cref{sec:method}, we selected two expert groups. 
For group (i), it was evident to invite the original authors of the concept of architectural perspectives. Additionally to their experience in industry, these experts can provide the necessary knowledge, history, and background on perspectives. While the decision on inviting the original authors can be considered as study design trade-off \cite{RobillardEtAl_CommunicatingStudy_2024}, it may also introduce bias and a potential threat to validity, discussed in \Cref{sec:threats}. For group (ii), we invited two industry experts in software architecture, capable of contributing knowledge on similar or alternative concepts. All experts have over 30 years of experience across diverse software domains and share a common understanding of software sustainability.

The focus group lasted one hour and was conducted virtually; it followed a protocol (see online material \cite{replication_package}) and was audio recorded. The post-process included transcribing, applying provisional coding, and synthesizing the results. The focus group has two goals: (i) determine whether the concept \textit{perspectives} is useful in practice, or whether the whole concept or only certain elements should be reconsidered to better meet industrial needs; and (ii) explore our experts' knowledge and industrial experience on sustainability to guide us in delivering a sustainability perspective relevant to current needs.
Below, we summarize the findings and phrase implications (\textbf{I-1} to \textbf{I-6}) for our research:
\\
\\
\noindent
\textbf{Synthesis.} We synthesis the implications according their emerged themes:
\\
\noindent
\textit{Perspectives in use.}
The participants noted that the perspective elements are no different from other concepts or techniques used in industry. All participants agreed that all concepts are either known in isolation or with a different name. For example, "for the activities, we often use the word process" (P3); or pitfalls are also called "trade-offs" (P4). %All participants agreed that no elements were least useful.
All participants agreed that no element could be considered unnecessary or without relevance.

The checklists were identified as a particularly valuable tool in high-pressure environments where architects may overlook important considerations (P2; P3). Concerns were raised, however, that checklists could "cause people to stop thinking" (P3), especially when used rigidly. P3 suggested that open-ended templates or forms would work better in an industrial context, as they trigger reflection and are therefore more effective.

While tactics are seen as important, all participants felt that they could quickly become outdated as technologies evolve. It was suggested that tactics might be decoupled from the perspectives in order to remain flexible: "I would certainly decouple as much as possible the tactics section from the other sections, which are much more future proof" (P3).

All participants agreed that there are no elements that are least useful, but that experts tend to skip over applicability and concerns in particular. "I would say this is a sign of immaturity. If people cannot state for a particular perspective when it is not applicable, it just means that it is not mature. Because they have not seen it go wrong" (P3).

\begin{tcolorbox}
\scriptsize
\textbf{I-1:} Activities might be rephrased as "Process" to better support generally known concepts. 
\\ \textbf{I-2:} Although checklists are considered one of the most valuable element, we might change them to templates with open questions to stimulate reflection and critical thinking.
\\ \textbf{I-3:} Tactics could not be directly included in perspectives, but could be linked to external archives that are already established and kept up to date. This would allow for a perspective that is company- or domain-agnostic. Companies are already starting to collect their own sustainability tactics that have proven helpful in their own context.
\end{tcolorbox}

\paragraph{Misinterpretation of perspectives.}
Originally, perspectives were developed to store recurring architectural knowledge in a knowledge base "which people used to go and manipulate their views in order to achieve quality attributes"; and to avoid "having multiple architectural descriptions" (P1). 
However, the perspective authors reflected that this concept is not generally understood. Instead, architects are tempted to use perspectives as a knowledge representation \cite{AliBabarEtAl_SoftwareArchitecture_2009} to structure their architecture documentation. 
"In every architectural description, there is a section on security, a section on performance and scalability, a section on resilience, and a section on evolution [...]. I have yet to meet anyone apart from [P2] or I, who picked that subtlety up" (P1).
This is supported by P4: "I have never used perspectives like this [...] but it seems like it is a very natural part of the [architecture] design process" (P4). The misinterpretation can also be found in Kiwelekar et al. \cite{KiwelekarEtAl_ArchitecturalPerspective_2020}, who are \textit{creating} the learning analytics perspective \textit{at} system design (see \Cref{sec:snowballing}).

When perspectives are considered as a knowledge base, participants (P1; P2) mentioned that perspectives can support self-study and education in the form of both guiding people into a new area and providing a memory aid for experienced architects. 
Using a sustainability perspective as knowledge base to create and manipulate views, also suggests the integration of perspectives into computer science education -- as outlined in our introduction -- to help shift the curriculum toward addressing sustainability concerns; a need also identified by Peters et al. \cite{PetersEtAl_SustainabilityComputing_2024}.

\begin{tcolorbox}
\scriptsize
\textbf{I-4}: Regardless of how perspectives are ultimately used in practice, the elements in a perspective are indeed helpful in addressing certain concerns. To provide architects with the knowledge about sustainability that is not yet fully explored in all domains, we put forward the educational aspects of perspectives (for both professional education and computer science education). Over time, we may find that our sustainability perspective naturally evolves from a knowledge base to a dedicated chapter in the architectural description (as proposed in \cite{FunkeLago_CarvingSustainability_2023}).
\end{tcolorbox}

\paragraph{Perspectives as roles.}
Rather than treating perspectives as an abstract concept, most participants (P1; P2; P3) concluded that assigning perspectives to specific roles would allow for better alignment with industry practice. In particular, security concerns are often already managed by designated roles such as the security officer within organizations. This approach ensures that specific concerns are systematically addressed by individuals responsible for those concerns: "what we suggest is that [organizations] have champions for different key quality attributes" (P1).

\begin{tcolorbox}
\scriptsize
\textbf{I-5:} Role-based assignments promote accountability by having designated individuals oversee the integration of key quality attributes. While conducting a Delphi study \cite{Schmidt_ManagingDelphi_1997} and focus group in research phase 2, we may also include sustainability officers or sustainability "champions" to drive sustainability individually forward.
\end{tcolorbox}

\paragraph{Sustainability as perspective.}
When we asked for recommendations on the design of a sustainability perspective, most participants (P1; P2; P3) pointed out the importance of concerns within a perspective. "I think the concerns for sustainability are the most important because [sustainability] is such an undefined concept and unknown [...] concerns help people think" (P4).

P1 stressed that due to the complexity of sustainability, it might be necessary to shape multiple perspectives: "Sustainability is an awfully big topic, especially with your very holistic view of it. You might end up with multiple perspectives [...] but start with one and see how complex it gets" (P1). This is also supported by the experience from P1 and P2 as they encountered the same issue while defining their own perspectives: "when we looked at performance [...] we went round in circles as to whether that was two perspectives or just one".

Three participants (P1; P2; P3) underlined that especially the activities, pitfalls, and checklists mostly emerge from past experience. P1 called this the "validation filter": collecting all potential experience gained from past projects, and then validating it through the literature---and vice versa. "It [is] a combination of what is the collective wisdom of the literature we can find. And then, what is our experience?" (P1).

\begin{tcolorbox}
\scriptsize
\textbf{I-6:} 
To define relevant concerns within a sustainability perspective, we combine expert insights on industry needs with our own research experience. This interaction supports the iterative development of our artifact and helps determine whether a single holistic or multiple sustainability perspectives are needed. As sustainability-related experience and knowledge is still limited, the initial knowledge must be drawn from expert judgment and literature, then results need to be validated by industrial or controlled experiments.
\end{tcolorbox}

%%%%%%%%%%%%%%%%%%%%%
%%% Our Vision
%%%%%%%%%%%%%%%%%%%%%
%\section{A Sustainability Perspective - Our Vision}
\section{Discussion}
\label{sec:vision}
In the following, we first frame the results from Sections \ref{sec:snowballing} and \ref{sec:focus-group} in our sustainability perspective vision. Then, we provide an illustrative example that helps understanding how such a perspective, once defined, could be contextualized during the architecture design process.

\begin{table}
    \scriptsize
    \caption{\color{black}Sustainability perspective vision.}
    \label{tab:sus-perspective}
    \renewcommand{\arraystretch}{1.2}
    \begin{tabularx}{\linewidth}{p{80pt}|X}
        \textbf{Elements as in \cite{RozanskiWoods_SoftwareSystems_2012}} & \textbf{Our vision grounded in linked implications} \\
        \toprule
        \cellcolor{gray!15} \textbf{Desired Quality}	  & \textit{"Sustainability" or a more fine-grained definition, based on the Delphi consensus in research phase 2.} \newline Link: \textbf{I-5; I-6} \\ \hline
        \cellcolor{gray!15} \textbf{Applicability}	  & \textit{Set of viewpoints, based on the Delphi consensus in research phase 2.} \newline Link: \textbf{I-5; I-6} \\ \hline
        \cellcolor{gray!15} \textbf{Concerns}	  & \textit{Set of concerns, based on the Delphi consensus in research phase 2.} \newline Link: \textbf{I-5; I-6} \\ \hline
        \cellcolor{gray!15} \textbf{Activities}	   & \textit{Rephrase to "process". Will require the most research in terms of analyzing existing and already validated results in literature and practice.} \newline Link: \textbf{I-1; I-5; I-6} \\ \hline
        \cellcolor{gray!15} \textbf{Architectural tactics}	 & \textit{Decouple and link to existing archives that have emerged in research; for a specific domain; or even in a particular organization.} \newline Link: \textbf{I-3} \\ \hline
        \cellcolor{gray!15} \textbf{Problems and pitfalls}	  & \textit{Experience-based list of common problems faced in addressing sustainability issues, supplemented by insights gained in research phase 2 and 3.} \newline Link: \textbf{I-5; I-6}  \\ \hline
        \cellcolor{gray!15} \textbf{Checklist}	  & \textit{Formulate templates with open-ended questions. Based on experience-based knowledge, as described for} Problems and pitfalls. \newline Link: \textbf{I-2; I-5; I-6}
        \\
        \bottomrule
    \end{tabularx}
\end{table}

\paragraph{Sustainability Perspective Vision.}
In \Cref{tab:sus-perspective}, we present our vision for a sustainability perspective based on the implications (\textbf{I-1} to \textbf{I-6}) and evidence derived so far. Rather than already defining and filling each element, we outline how we intend to \textit{design} them. 
The actual content of the perspective will then be largely influenced by the results of our upcoming artifact design and validation cycles. 

Naturally, the \textbf{desired quality} is sustainability; however, we remain open to adjusting the focus based on the results of the Delphi method \cite{Schmidt_ManagingDelphi_1997}. This will help us decide whether the perspective should focus holistically on sustainability or whether we need to break it down into smaller concerns (e.g., environmental concerns, energy consumption concerns) leading to multiple sustainability perspectives. Our Delphi will also determine which viewpoints the perspective can be \textbf{applied} to, and which quality \textbf{concerns} the perspective should address.

Most of our research effort will focus on designing those \textbf{activities} which provide actual guidance. The activities will guide architects in (i) revising or creating views with sustainability in mind, (ii) applying the tactics, and (iii) anticipating potential challenges. While we will draw on existing research from Jagroep et al. \cite{JagroepEtAl_EnergyConsumption_2015} on their energy efficiency perspective, we aim to incorporate more recent findings. For example, we plan to incorporate a workflow for measuring and applying architectural tactics \cite{FunkeEtAl_ExperimentalEvaluation_2024}, a systematic method for measuring energy hotspots \cite{GuldnerEtAl_DevelopmentEvaluation_2024}, and a toolkit for designing sustainability-aware software at the architectural level \cite{LagoEtAl_SustainabilityAssessmentToolkit_2024}. The specific activities will depend on the particular concerns and perspectives.

We will follow the advice of our experts and decouple the \textbf{tactics} from the perspective to prevent them from becoming obsolete and offering a more flexible perspective meeting the needs of today's dynamic landscape. As the tactics are closely related to the specified quality concerns, they will also build on previous research, such as the categorization for energy efficiency tactics by Vos et al. \cite{VosEtAl_ArchitecturalTactics_2022} and others.

To formulate the \textbf{problems and pitfalls}, we need to draw on existing experiences as defined in the literature and supplement them with insights from our artifact validation and evaluation in research phases 2 and 3. The problems and pitfalls are likely to evolve over time as the sustainability perspective is applied more frequently to real software systems, so that we can learn from things that go wrong.

We aim to keep the \textbf{checklists}, but shaping them in the form of open questions as suggested. This will stimulate critical thinking. The questions themselves will also be derived from experience and will probably be extended over time.

\begin{figure}
    \centering
    \includegraphics[width=0.6\linewidth]{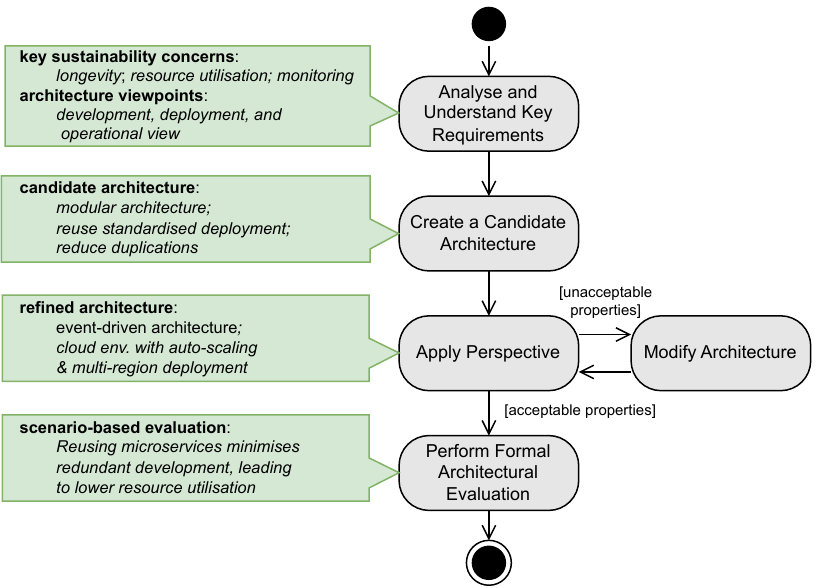}
    \caption{Perspective usage reused from \cite{WoodsRozanski_UsingArchitectural_2005} and complemented with example outputs (green callouts) concerning sustainability.}
    \label{fig:example}
\end{figure}

\paragraph{Sustainability Perspective in Action.}
\label{sec:example}
\Cref{fig:example} provides a tangible example including outcomes. We reuse the UML activity diagram and description proposed by Woods and Rozanski \cite{WoodsRozanski_UsingArchitectural_2005} and complement them with an illustrative example for sustainability. The outcomes are only exemplary and solely based on our own experience -- they are not derived from a real scenario or case. Filling this process with data will be part of our iterative artifact design and the last research phase 3.

Based on the requirements, architects first identify \textbf{key sustainability concerns} and relevant \textbf{viewpoints}, either from the literature (e.g., 4+1 viewpoints \cite{Kruchten_4+1View_1995}, catalog of viewpoints \cite{RozanskiWoods_SoftwareSystems_2012}), or from industry (e.g., TOGAF \cite{TheOpenGroup_TOGAFStandard_2022}).
In our example, potential key sustainability concerns might be (i) \textit{"use modular and maintainable code to extend system longevity"}, (ii) \textit{"optimize infrastructure to reduce resource waste"}, and (iii) \textit{"implement monitoring tools to monitor sustainability incentives"}. These concerns would imply the usage of a (i) \textit{development view}, (ii) \textit{deployment view}, and (iii) \textit{operational view}; each targeting one of the specified sustainability concerns accordingly.

The architect designs an initial \textbf{candidate architecture}. Our concerns suggest a modular architecture as architecture style and reusable deployment components as technology stack to reduce duplications. As emphasized by Woods and Rozanski \cite{WoodsRozanski_UsingArchitectural_2005}, the step of creating a candidate architecture and applying the perspective is usually performed together and not in sequence or isolation. 

For each identified sustainability concern, we now apply the sustainability perspective\footnote{\scriptsize A perspective can only be effectively applied if it was designed to address the intended concerns.} leading to a \textbf{refined architecture}. In our case, we might use available frameworks for architecture design addressing sustainability concerns and identify architecture tactics accordingly. For instance, potential frameworks from the literature might lead us to tactics suggesting a containerized microservice architecture ensuring reusability, scalability, and suitability for cloud environments (c.f. our perspective vision above for potential frameworks and tactic archives targeting relevant sustainability concerns). 

An additional variability feature analysis \cite{FunkeEtAl_VariabilityFeatures_2023} might reveal that serverless functions can reduce idle compute resources. Open-ended questions within our perspective would stimulate further reflection, such as \textit{"How are cloud resources allocated to minimize resource allocation and support scalability?"}. The process ends with a formal architecture evaluation, such as \textbf{scenario-based evaluation} ensuring that the defined sustainability concerns are met.

In summary, a perspective is not a single rigid recipe, rather a complete, living cookbook filled with several recipes that one can reuse. However, as emphasized by Woods and Rozanski as well as by our experts in the focus group, applying a perspective does not spare the architect a trade-off analysis \cite{WoodsRozanski_UsingArchitectural_2005}.

%%%%%%%%%%%%%%%%%%%%%
%%% Threats
%%%%%%%%%%%%%%%%%%%%%
\color{black}
\section{Threats to Validity}
\label{sec:threats}
As in all studies, also this research has potential threats to validity. We focus on threats related to our present executed work, i.e., the literature review and focus group. We follow the classification according to Wohlin et al. \cite{WohlinEtAl_ExperimentationSoftware_2012}.

\textbf{Literature Review.}
The main threat in our current research is related to the sole use of forward snowballing to identify related studies and is hence classified as potential threat to \textit{reliability}. However, research has been successfully used forward snowballing as alternative (see our snowballing description in \Cref{sec:method}) and our results provided valuable related literature around architecture perspectives and perspectives for sustainability. To further mitigate the threat of missing out relevant studies, we (i) performed one additional round of backward snowballing on our primary studies, and (ii) executed one additional round of forward snowballing on the most related primary study to ours, i.e., the work from Jagroep~et~al.~\cite{JagroepEtAl_ExtendingSoftware_2017}.

With regard to \textit{conclusion validity}, a potential selection bias occurred as our forwarding snowballing seed is only based on the two seminal works from Woods and Rozanski. While there might be indeed related concepts available, we are particularly interested in architectural perspectives according to their definition. This allows us to evaluate perspectives in a new light, and facilitate them for sustainability. We mitigate this threat by our subsequent focus group, questioning experts about related concepts.

\textbf{Focus Group.}
One threat in relation to our focus group regards \textit{external validity} and concerns sampling bias. Indeed, based on four experts, we cannot generalize their opinions to the whole population of all architects. However, as we did not aim at generalizability but rather at significance, we can rely on their industrial experience of more than 30 years on average.

A second threat concerns \textit{construct validity}. As we have invited the original authors of perspectives, we might introduce biased responses as their involvement could influence other participants' opinion. However, at this point we had to make a study design trade-off \cite{RobillardEtAl_CommunicatingStudy_2024} between inviting experts who have actually applied perspectives in their daily work -- or not. We chose the former to gain more reliable insights. We mitigated this threat by having also two experts on the counter part who have not used perspectives.

%%%%%%%%%%%%%%%%%%%%%
%%% Conclusion
%%%%%%%%%%%%%%%%%%%%%
\section{Conclusion and Future Plans}
\label{sec:conclusion}
In this work, we contribute what we call a \textit{sustainability perspective vision}, i.e., a revised notion of architectural perspective (based on the literature and experience in the field) meant to be filled with the own definition of sustainability.
Starting from these initial results, we will continue with research Phase 2 by building a concrete panel for our Delphi and executing it to derive the necessary sustainability concerns our perspective should focus on. Then, we will design the concrete activities that target exactly such derived sustainability concerns coming from the Delphi. We need to ensure that the activities are both actionable and relevant to real-world scenarios. Finally, in research Phase 3, we need to evaluate the perspective in a multi-case study. The case studies will also provide us with valuable insights on how perspectives can be integrated into the current way of working (refining our example in \Cref{sec:example}), such as in combination with established frameworks or tools (e.g., using architecture views and viewpoints in TOGAF \cite{TheOpenGroup_TOGAFStandard_2022}).

\begin{credits}
\scriptsize\subsubsection{\scriptsize\ackname} 
This project is partially supported by (i) the LETSGO Project, promoted by the Netherlands Enterprise Agency (Rijksdienst voor Ondernemend Nederland) and (ii) the Sustainable IT - Lab of the Vrije Universiteit Amsterdam and ABN AMRO Bank N.V. We thank our focus group members Grace Lewis, Eltjo Poort, Nick Rozanski, and Eoin Woods.

% \subsubsection{\discintname}
% It is now necessary to declare any competing interests or to specifically
% state that the authors have no competing interests. Please place the
% statement with a bold run-in heading in small font size beneath the
% (optional) acknowledgments\footnote{If EquinOCS, our proceedings submission
% system, is used, then the disclaimer can be provided directly in the system.},
% for example: The authors have no competing interests to declare that are
% relevant to the content of this article. Or: Author A has received research
% grants from Company W. Author B has received a speaker honorarium from
% Company X and owns stock in Company Y. Author C is a member of committee Z.
\end{credits}
\bibliographystyle{splncs04}
\bibliography{03_references}

\begin{thebibliography}{10}
\providecommand{\url}[1]{\texttt{#1}}
\providecommand{\urlprefix}{URL }
\providecommand{\doi}[1]{https://doi.org/#1}

\bibitem{AliBabarEtAl_SoftwareArchitecture_2009}
Ali~Babar, M., Dings{\o}yr, T., Lago, P., {van Vliet}, H.: Software {{Architecture Knowledge Management}}. Springer (2009)

\bibitem{AndrikopoulosEtAl_SustainabilitySoftware_2022}
Andrikopoulos, V., Boza, R.D., Perales, C., Lago, P.: Sustainability in {{Software Architecture}}: {{A Systematic Mapping Study}}. In: 2022 48th {{Euromicro Conference}} on {{Software Engineering}} and {{Advanced Applications}} ({{SEAA}}). IEEE (2022)

\bibitem{BassEtAl_SoftwareArchitecture_2021}
Bass, L., Clements, P., Kazman, R.: Software Architecture in Practice. Sei Series in Software Engineering, Addison-Wesley, fourth edition edn. (2021)

\bibitem{BeckerEtAl_SustainabilityDesign_2015}
Becker, C., Chitchyan, R., Duboc, L., Easterbrook, S., Penzenstadler, B., Seyff, N., Venters, C.C.: Sustainability {{Design}} and {{Software}}: {{The Karlskrona Manifesto}}. In: {{37th International Conference}} on {{Software Engineering}} (2015)

\bibitem{BouckeHolvoet_RelatingArchitectural_2006}
Bouck{\'e}, N., Holvoet, T.: Relating architectural views with architectural concerns. In: International Workshop on {{Early}} Aspects at {{ICSE}}. ACM (2006)

\bibitem{CaleroEtAl_5WsGreen_2020}
Calero, C., Mancebo, J., Garcia, F., Moraga, M.A., Berna, J.A.G., {Fernandez-Aleman}, J.L., Toval, A.: {{5Ws}} of green and sustainable software. Tsinghua Science and Technology  \textbf{25}(3) (2020)

\bibitem{CapillaEtAl_10Years_2016}
Capilla, R., Jansen, A., Tang, A., Avgeriou, P., Babar, M.A.: 10 years of software architecture knowledge management: {{Practice}} and future. Journal of Systems and Software  \textbf{116} (2016)

\bibitem{ChitchyanEtAl_SustainabilityDesign_2016}
Chitchyan, R., Becker, C., Betz, S., Duboc, L., Penzenstadler, B., Seyff, N., Venters, C.C.: Sustainability design in requirements engineering: State of practice. In: {{38th International Conference}} on {{Software Engineering Companion}}. ACM (2016)

\bibitem{DingaEtAl_EmpiricalEvaluation_2023}
Dinga, M., Malavolta, I., Giamattei, L., Guerriero, A., Pietrantuono, R.: An {{Empirical Evaluation}} of the {{Energy}} and {{Performance Overhead}} of {{Monitoring Tools}} on {{Docker-Based Systems}}. In: Service-{{Oriented Computing}}, vol. 14419. Springer Nature Switzerland (2023)

\bibitem{FelizardoEtAl_UsingForward_2016}
Felizardo, K.R., Mendes, E., Kalinowski, M., Souza, {\'E}.F., Vijaykumar, N.L.: Using {{Forward Snowballing}} to update {{Systematic Reviews}} in {{Software Engineering}}. In: {{10th International Symposium}} on {{Empirical Software Engineering}} and {{Measurement}}. ACM (2016)

\bibitem{FunkeLago_CarvingSustainability_2023}
Funke, M., Lago, P.: Carving {{Sustainability}} into {{Architecture Knowledge Practice}}. In: Software {{Architecture}}. Lecture {{Notes}} in {{Computer Science}}, vol. 14212. Springer Nature Switzerland (2023)

\bibitem{poster_FunkeLago_2024}
Funke, M., Lago, P.: Towards an {{Architectural Perspective}} for {{Sustainability}}. In: Companion {{Proceedings}} of the 15th {{International Conference}} on {{Software Business}}. {{CEUR Workshop Proceedings}}, vol.~3921 (2024)

\bibitem{replication_package}
Funke, M., Lago, P.: {Replication Package} (2025), \url{https://github.com/S2-group/QUATIC-2025-sustainability-perspective-rep-pkg}

\bibitem{FunkeEtAl_ExperimentalEvaluation_2024}
Funke, M., Lago, P., Adenekan, E., Malavolta, I., Heitlager, I.: Experimental {{Evaluation}} of {{Energy Efficiency Tactics}} in {{Industry}}: {{Results}} and {{Lessons Learned}}. In: {{International Conference}} on {{Software Architecture}} ({{ICSA}}). IEEE (2024)

\bibitem{FunkeEtAl_VariabilityFeatures_2023}
Funke, M., Lago, P., Verdecchia, R.: Variability {{Features}}: {{Extending Sustainability Decision Maps}} via an {{Industrial Case Study}}. In: 20th {{International Conference}} on {{Software Architecture Companion}} ({{ICSA-C}}). IEEE (2023)

\bibitem{GuldnerEtAl_DevelopmentEvaluation_2024}
Guldner, A., et~al.: Development and evaluation of a reference measurement model for assessing the resource and energy efficiency of software products and components---{{Green Software Measurement Model}} ({{GSMM}}). Future Generation Computer Systems  \textbf{155} (2024)

\bibitem{GurbuzEtAl_SafetyPerspective_2014}
G{\"u}rb{\"u}z, H.G., Tekinerdogan, B., Pala~Er, N.: Safety {{Perspective}} for {{Supporting Architectural Design}} of {{Safety-Critical Systems}}. In: Software {{Architecture}}, vol.~8627. Springer International Publishing (2014)

\bibitem{HeldalEtAl_SustainabilityCompetencies_2024}
Heldal, R., et~al.: Sustainability competencies and skills in software engineering: {{An}} industry perspective. Journal of Systems and Software  \textbf{211} (2024)

\bibitem{ISO_42010_2022}
{International Organization for Standardization [ISO]}: Systems and software engineering -- {{Architecture}} description. Tech. Rep. ISO/IEC/IEEE 42010:2022 (2022)

\bibitem{JagroepEtAl_ExtendingSoftware_2017}
Jagroep, E., Van Der~Werf, J.M., Brinkkemper, S., Blom, L., Van~Vliet, R.: Extending software architecture views with an energy consumption perspective: {{A}} case study on resource consumption of enterprise software. Computing  \textbf{99}(6) (2017)

\bibitem{JagroepEtAl_EnergyConsumption_2015}
Jagroep, E.A., Van Der~Werf, J.M.E.M., Spauwen, R., Blom, L., Van~Vliet, R., Brinkkemper, S.: An {{Energy Consumption Perspective}} on {{Software Architecture}}. In: Software {{Architecture}}, vol.~9278. Springer (2015)

\bibitem{KanelVecchiola_GlobalTechnology_2013}
K{\"a}nel, J.V., Vecchiola, C.: Global technology trends: Perspectives from {{IBM Research Australia}} on resilient systems. International Journal of Computational Science and Engineering  \textbf{8}(3) (2013)

\bibitem{KiczalesEtAl_AspectorientedProgramming_1997}
Kiczales, G., Lamping, J., Mendhekar, A., Maeda, C., Lopes, C., Loingtier, J.M., Irwin, J.: Aspect-oriented programming. In: {{ECOOP}}'97 --- {{Object-Oriented Programming}}, vol.~1241. Springer Berlin Heidelberg (1997)

\bibitem{KiwelekarEtAl_ArchitecturalPerspective_2020}
Kiwelekar, A.W., Laddha, M.D., Netak, L.D., Gandhi, S.: An {{Architectural Perspective}} of {{Learning Analytics}}. In: Machine {{Learning Paradigms}}, vol.~158. Springer International Publishing (2020)

\bibitem{KontioEtAl_FocusGroup_2008}
Kontio, J., Bragge, J., Lehtola, L.: The {{Focus Group Method}} as an {{Empirical Tool}} in {{Software Engineering}}. In: Guide to {{Advanced Empirical Software Engineering}}. Springer London (2008)

\bibitem{KoziolekEtAl_MORPHOSISLightweight_2012a}
Koziolek, H., Domis, D., Goldschmidt, T., Vorst, P., Weiss, R.J.: {{MORPHOSIS}}: {{A Lightweight Method Facilitating Sustainable Software Architectures}}. In: 2012 {{Joint Working IEEE}}/{{IFIP Conference}} on {{Software Architecture}} and {{European Conference}} on {{Software Architecture}}. IEEE (2012)

\bibitem{Kruchten_4+1View_1995}
Kruchten, P.: The 4+1 {{View Model}} of architecture. IEEE Software  \textbf{12}(6) (1995)

\bibitem{LagoEtAl_SustainabilityAssessmentToolkit_2024}
Lago, P., Condori~Fernandez, N., Fatima, I., Funke, M., Malavolta, I.: The sustainability assessment framework toolkit: A decade of modeling experience. Software and Systems Modeling  (2024)

\bibitem{LagoEtAl_FramingSustainability_2015}
Lago, P., Ko{\c c}ak, S.A., Crnkovic, I., Penzenstadler, B.: Framing sustainability as a property of software quality. Communications of the ACM  \textbf{58}(10) (2015)

\bibitem{LwakatareEtAl_LargescaleMachine_2020}
Lwakatare, L.E., Raj, A., Crnkovic, I., Bosch, J., Olsson, H.H.: Large-scale machine learning systems in real-world industrial settings: {{A}} review of challenges and solutions. Information and Software Technology  \textbf{127} (2020)

\bibitem{MahauxEtAl_DiscoveringSustainability_2011}
Mahaux, M., Heymans, P., Saval, G.: Discovering {{Sustainability Requirements}}: {{An Experience Report}}. In: Requirements {{Engineering}}: {{Foundation}} for {{Software Quality}}, vol.~6606. Springer Berlin Heidelberg (2011)

\bibitem{Martin-MartinEtAl_GoogleScholar_2018}
{Mart{\'i}n-Mart{\'i}n}, A., {Orduna-Malea}, E., Thelwall, M., {Delgado L{\'o}pez-C{\'o}zar}, E.: Google {{Scholar}}, {{Web}} of {{Science}}, and {{Scopus}}: {{A}} systematic comparison of citations in 252 subject categories. Journal of Informetrics  \textbf{12}(4) (2018)

\bibitem{McGuireEtAl_SustainabilityStratified_2023}
McGuire, S., Schultz, E., Ayoola, B., Ralph, P.: Sustainability is {{Stratified}}: {{Toward}} a {{Better Theory}} of {{Sustainable Software Engineering}}. In: {{International Conference}} on {{Software Engineering}} ({{ICSE}}). IEEE (2023)

\bibitem{PathaniaEtAl_KnowledgeBase_2023}
Pathania, P., Mehra, R., Sharma, V.S., Kaulgud, V., Podder, S., Burden, A.P.: Towards a {{Knowledge Base}} of {{Common Sustainability Weaknesses}} in {{Green Software Development}}. In: {{38th International Conference}} on {{Automated Software Engineering}} ({{ASE}}). IEEE (2023)

\bibitem{PereiraEtAl_RankingProgramming_2021}
Pereira, R., Couto, M., Ribeiro, F., Rua, R., Cunha, J., Fernandes, J.P., Saraiva, J.: Ranking programming languages by energy efficiency. Science of Computer Programming  \textbf{205} (2021)

\bibitem{PetersEtAl_SustainabilityComputing_2024}
Peters, A.K., et~al.: Sustainability in {Computing} {Education}: {A} {Systematic} {Literature} {Review}. ACM Transactions on Computing Education  (2024)

\bibitem{ProcacciantiEtAl_GreenLab_2015}
Procaccianti, G., Lago, P., Vetro, A., Fernandez, D.M., Wieringa, R.: The {{Green Lab}}: {{Experimentation}} in {{Software Energy Efficiency}}. In: {{International Conference}} on {{Software Engineering}}. IEEE (2015)

\bibitem{RaniEtAl_EnergyPatterns_2024}
Rani, P., Zellweger, J., Kousadianos, V., Cruz, L., Kehrer, T., Bacchelli, A.: Energy {{Patterns}} for {{Web}}: {{An Exploratory Study}}. In: Proceedings of the 46th {{International Conference}} on {{Software Engineering}}: {{Software Engineering}} in {{Society}}. ACM (2024)

\bibitem{RobillardEtAl_CommunicatingStudy_2024}
Robillard, M.P., Arya, D.M., Ernst, N.A., Guo, J.L., Lamothe, M., Nassif, M., Novielli, N., Serebrenik, A., Steinmacher, I., Stol, K.J.: Communicating {{Study Design Trade-offs}} in {{Software Engineering}}. ACM Transactions on Software Engineering and Methodology  (2024)

\bibitem{RozanskiWoods_SoftwareSystems_2012}
Rozanski, N., Woods, E.: Software Systems Architecture: Working with Stakeholders Using Viewpoints and Perspectives. Addison-Wesley, 2nd edn. (2012)

\bibitem{Ruparelia_SoftwareDevelopment_2010}
Ruparelia, N.B.: Software development lifecycle models. ACM SIGSOFT Software Engineering Notes  \textbf{35}(3) (2010)

\bibitem{SantAnnaEtAl_MasteringCrosscutting_2013}
Sant'Anna, C., Garcia, A., Batista, T., Rashid, A.: Mastering crosscutting architectural decisions with aspects. Software: Practice and Experience  \textbf{43}(3) (2013)

\bibitem{Schmidt_ManagingDelphi_1997}
Schmidt, R.C.: Managing {{Delphi Surveys Using Nonparametric Statistical Techniques}}. Decision Sciences  \textbf{28}(3) (1997)

\bibitem{TangEtAl_ComparativeStudy_2010}
Tang, A., Avgeriou, P., Jansen, A., Capilla, R., Ali~Babar, M.: A comparative study of architecture knowledge management tools. Journal of Systems and Software  \textbf{83}(3) (2010)

\bibitem{TekinerdoganOzcan_ArchitecturalPerspective_2017}
Tekinerdogan, B., Ozcan, O.: Architectural {{Perspective}} for {{Design}} and {{Analysis}} of {{Scalable Software}} as a {{Service Architectures}}. In: Managing {{Trade-Offs}} in {{Adaptable Software Architectures}}. Elsevier (2017)

\bibitem{TheOpenGroup_TOGAFStandard_2022}
{The Open Group}: The {{TOGAF}}{\textregistered} {{Standard}}, {{Version}} 10. Tech. rep. (2022), \url{https://pubs.opengroup.org/architecture/w212/#_ftn1}

\bibitem{VentersEtAl_SustainableSoftware_2023}
Venters, C.C., Capilla, R., Nakagawa, E.Y., Betz, S., Penzenstadler, B., Crick, T., Brooks, I.: Sustainable software engineering: {{Reflections}} on advances in research and practice. Information and Software Technology  (2023)

\bibitem{VolpatoEtAl_HasSocial_2019}
Volpato, T., Allian, A., Nakagawa, E.Y.: Has social sustainability been addressed in software architectures? In: Proceedings of the 13th {{European Conference}} on {{Software Architecture}} - {{Volume}} 2. ACM (2019)

\bibitem{VosEtAl_ArchitecturalTactics_2022}
Vos, S., Lago, P., Verdecchia, R., Heitlager, I.: Architectural {{Tactics}} to {{Optimize Software}} for {{Energy Efficiency}} in the {{Public Cloud}}. In: {ICT4S}. IEEE (2022)

\bibitem{Wieringa_DesignScience_2014}
Wieringa, R.J.: Design {{Science Methodology}} for {{Information Systems}} and {{Software Engineering}}. Springer (2014)

\bibitem{Wohlin_GuidelinesSnowballing_2014}
Wohlin, C.: Guidelines for snowballing in systematic literature studies and a replication in software engineering. In: Proceedings of the 18th {{International Conference}} on {{Evaluation}} and {{Assessment}} in {{Software Engineering}}. ACM (2014)

\bibitem{WohlinEtAl_SuccessfulCombination_2022}
Wohlin, C., Kalinowski, M., Romero~Felizardo, K., Mendes, E.: Successful combination of database search and snowballing for identification of primary studies in systematic literature studies. Information and Software Technology  \textbf{147} (2022)

\bibitem{WohlinEtAl_ExperimentationSoftware_2012}
Wohlin, C., Runeson, P., Höst, M., Ohlsson, M.C., Regnell, B., Wesslén, A.: Experimentation in {Software} {Engineering}. Springer Berlin Heidelberg (2012)

\bibitem{WoodsRozanski_UsingArchitectural_2005}
Woods, E., Rozanski, N.: Using {{Architectural Perspectives}}. In: 5th {{Working IEEE}}/{{IFIP Conference}} on {{Software Architecture}} ({{WICSA}}'05) (2005)

\end{thebibliography}

\end{document}